\documentclass[aps,prl,twocolumn,groupedaddress,showpacs]{revtex4-1}

\usepackage{graphicx}
\graphicspath{{./}}
\usepackage{caption}
\usepackage{subcaption}

\begin{document}


\title{Coexistence of multiple metastable polytypes in rhombohedral bismuth}



\author{Yu Shu$^{1}$, Wentao Hu$^{1}$, Zhongyuan Liu$^{1}$, Guoyin Shen$^{2}$,
Zhisheng Zhao$^{1}$, \\Julong He$^{1}$, Yanbin Wang$^{3}$, Yongjun Tian$^{1}$, Dongli Yu$^{1,}$
}
\email[]{ydl@ysu.edu.cn}
\affiliation{$^{1}$State Key Laboratory of Metastable Materials Science and Technology, Yanshan University, Qinhuangdao, Hebei 066004, China\\
$^{2}$High Pressure Collaborative Access Team, Geophysical Laboratory, Carnegie Institution of Washington, Argonne, IL 604390, USA\\
$^{3}$Center for Advanced Radiation Sources, University of Chicago, Chicago, Illinois 60439, USA
}

\date{\today}

\begin{abstract}
Derivative structural polytypes coexisting with the rhombohedral A7 structure of elemental bismuth (Bi) have been discovered at ambient condition, based on microstructure analyses of pure Bi samples treated under high pressure and high temperature conditions. Three structures with atomic positions close to those of the A7 structure have been identified through first-principles calculations, showing these polytypes energetically comparable to the A7 structure under ambient condition. Simulated diffraction data are in excellent agreement with the experimental observations. We argue that previously reported variations in physical properties (e.g., density, melting point, electrical conductivity, and magnetism) in bismuth could be due to the formation of these polytypes. The coexistence of metastable derivative structural polytypes may be a widely occurring phenomenon in other elemental materials.

\end{abstract}


\maketitle


Bismuth as a typical semimetal has attracted extensive research interests. Numerous studies have been conducted on Bi to investigate the electrical properties \cite{[]Smith1,Mangez2,Kope3,Behnia4,Hofmann5,Garcia6,Rogacheva7}, thermoelectric properties \cite{[]Heremans8,Hostler9}, and phase transitions under high pressure and high temperature (HP-HT) \cite{[]Tonkov10,Rimas11,Kabalkina12,Lu13,Chen14}, etc.. Under ambient condition, Bi is commonly designated as a rhombohedral lattice (space group R-3m, Strukturbericht A7), which is characterized by a pair of atoms spaced non-equidistantly along the trigonal axis in a Peierls distortion of the simple cubic structure \cite{[]Fritz16}. Alternatively, the structure can be described either as a hexagonal lattice with six atoms per unit cell or as a pseudo-cubic lattice with one atom per unit cell \cite{[]Hofmann5}. However, it is well-known that the observed X-ray diffraction (XRD) patterns of Bi cannot be fully accounted for with the rhombohedral A7 structure, which is designated as (marked by) ``Doubtful (?)" in the PDF card (JCPDS PDF $\#$44-1246) of the International Center for Diffraction Data and labeled with ``$\S$" in the periodic table of elements \cite{[]net15}. The symbol of ``$\S$" indicates crystal structure is unusual or may require further explanation.

Some long-standing unusual behaviors in physical properties have also been noted in Bi, possibly resulting from the structure uncertainty. In 1928, Kapitza reported inconsistent changes in specific resistance when Bi crystals were subjected to strong magnetic field \cite{[]Kapitza17}, and attributed the observations to small lattice distortions in Bi structure. Later, slight density variations of Bi single crystals were reported \cite{[]Goetz18}. In 1931, Goetz and Focke found that Bi single crystals grown under a strong oriented magnetic field exhibit a decrease in density and an increase in specific resistance \cite{[]Goetz19}. The applied magnetic field appeared to affect the Bi lattice during solidification, leading to changes in the density and resistance. All these earlier observations were attributed to a so-called ``mosaic block" structure \cite{[]Zwicky20,Zwicky21} with slightly different lattice constants \cite{[]Goetz18}. A variety of formation mechanisms were proposed for this mosaic structure based on different assumptions, such as the influence of impurities, allotropic changes, and deformation twinning \cite{[]Goetz18,Goetz22}. Nonetheless, the atomic configuration of the mosaic structure in Bi crystals remains elusive.

In this paper, we report structural polytypes of Bi evidenced by both experimental observations and first-principles calculations. We performed structural characterizations on pure Bi samples treated at various pressure and temperature conditions with XRD and high-resolution transmission electron microscopy (HRTEM) and discovered a remarkable structural diversity in Bi after the treatments. Using crystal structure prediction techniques, we identified three new structures of bismuth that are closely related to the well-known A7 structure and differ from each other only by minute lattice distortions. First-principles calculations indicate these structures have very similar energies, explaining the observed coexistence of multiple metastable polytypes in Bi. Such a structural diversity is intrinsic, not due to extrinsic action from the impurities or deformation, and leads to the property variations (such as density, melting point, electrical conductivity, and magnetism) observed in Bi in earlier studies. These polytypes provide a structural interpretation for the so-called ``mosaic block" structure previously speculated.

Bi granules (purity 99.999$\%$, Alfa Aesar) were compressed into cylinders ($\o$6mm$\times$6mm ) and subjected to HP-HT treatments in a six-anvil high pressure apparatus \cite{[]Liu23}. The HP-HT conditions are presented in \cite{[]Supplemental24}(Table SI). After the HP-HT treatments, the samples were quenched to room temperature before the pressure was unloaded. Both the raw granules and the bulk samples after HP-HT treatment were ground into powders for XRD (Rigaku D/MAX-PC/2500, Cu-K$\alpha$) and transmission electron microscopy (JEM-2010) characterizations.

Searches for energetically and dynamically possible structures of Bi were carried out at pressures from 0 to 6GPa with simulation cells consisting of 2--12 atoms using CALYPSO, without employing any known structure information \cite{[]Ma25}. Structural relaxation was then performed using the density functional theory (DFT) within the generalized gradient approximation (GGA)\cite{[]Perdew26}implemented in VASP \cite{[]Kresse27}. Phonon frequencies were generated using the CASTEP with finite displacement theory \cite{[]Clark28}. The calculation parameters are provided in \cite{[]Supplemental24}.It is known that conventional GGA calculations tend to overestimate the lattice parameters \cite{[]Philipp29}; hence a hydrostatic pressure of 1.2 GPa was applied to the A7 structure during structural relation, producing lattice parameters consistent with the experimentally determined ones under ambient condition (PDF $\#$44-1246). This pressure (1.2 GPa) was thus applied to all proposed structures in first-principles calculations.

An HRTEM image of the sample treated at 2 GPa and 2273 K viewed along [010] zone axis is presented in Fig. 1a with the corresponding selected area electron diffraction (SAED) patterns shown in Fig. 1b. Simulated atomic distribution of the ideal A7-Bi structure viewed along same zone axis is shown in Fig. 1c. In Fig. 1a, several regions with distinct atomic configurations are highlighted (yellow circles, labeled A--F), which are clearly different from the ideal A7 atomic distribution (Fig. 1c). The boundaries between these lattice distortion regions (Fig. 1a A--F) are fuzzy and poorly defined. For clarity, these regions are displayed side-by-side in Fig. 2, and the d-spacings and interplanar angles measured from these regions are summarized in Table SII [24]. All these regions show very similar d-spacings and interplanar angles, with deviations smaller than 0.1 \AA\ and 0.3$^o$, respectively. We note that the d-spacings ( 3.74$\pm$0.03 \AA\ and 3.97$\pm$0.03 \AA\ ) and the interplanar angles ( 71.52$\pm$0.05$^o$ ) determined from region E perfectly match those of the (003) and (101) planes of the A7 structure. Therefore, the SAED pattern (Fig. 1b) taken from the entire area of Fig. 1a does not show clear evidence of structural variation, even though changes in lattice parameters and clearly differences in lattice fringes are observed.
\begin{figure}[h]
\centering
\includegraphics[scale=0.16]{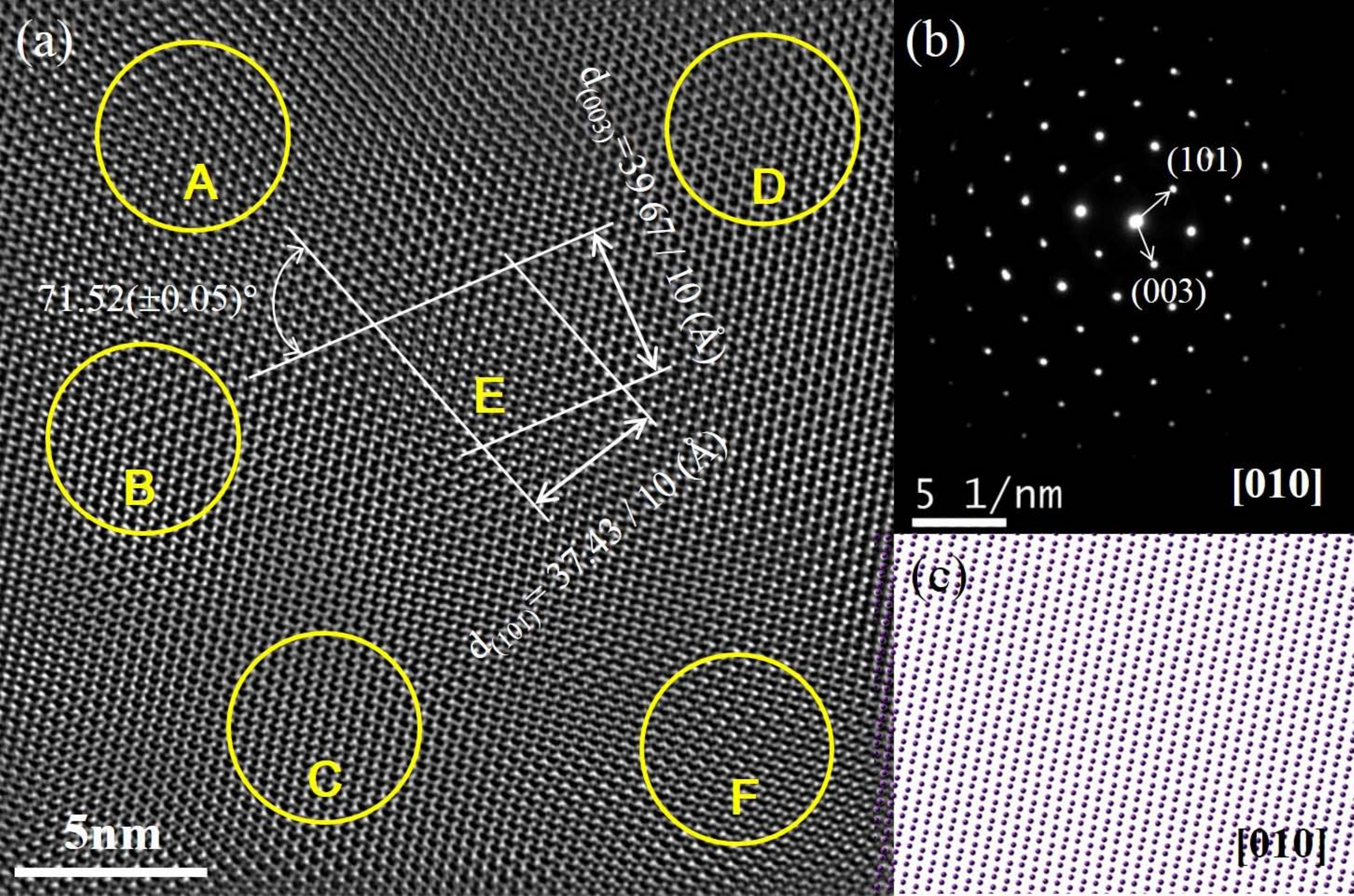}
\caption{\label{}TEM of Bi treated under 2GPa and 2273 K. (a) HRTEM image viewed along the [101] zone axis of the A7 structure. Different lattice distortion regions are labeled by A--F. Measurements of d-spacings and interplanar angles for the A7 structure are shown in region E; (b) Overall SEAD pattern corresponding to (a); (c) Simulated atomic distribution observed along [010] direction in A7-Bi structure.}
\end{figure}

\begin{figure}[h]
\centering
\includegraphics[scale=0.16]{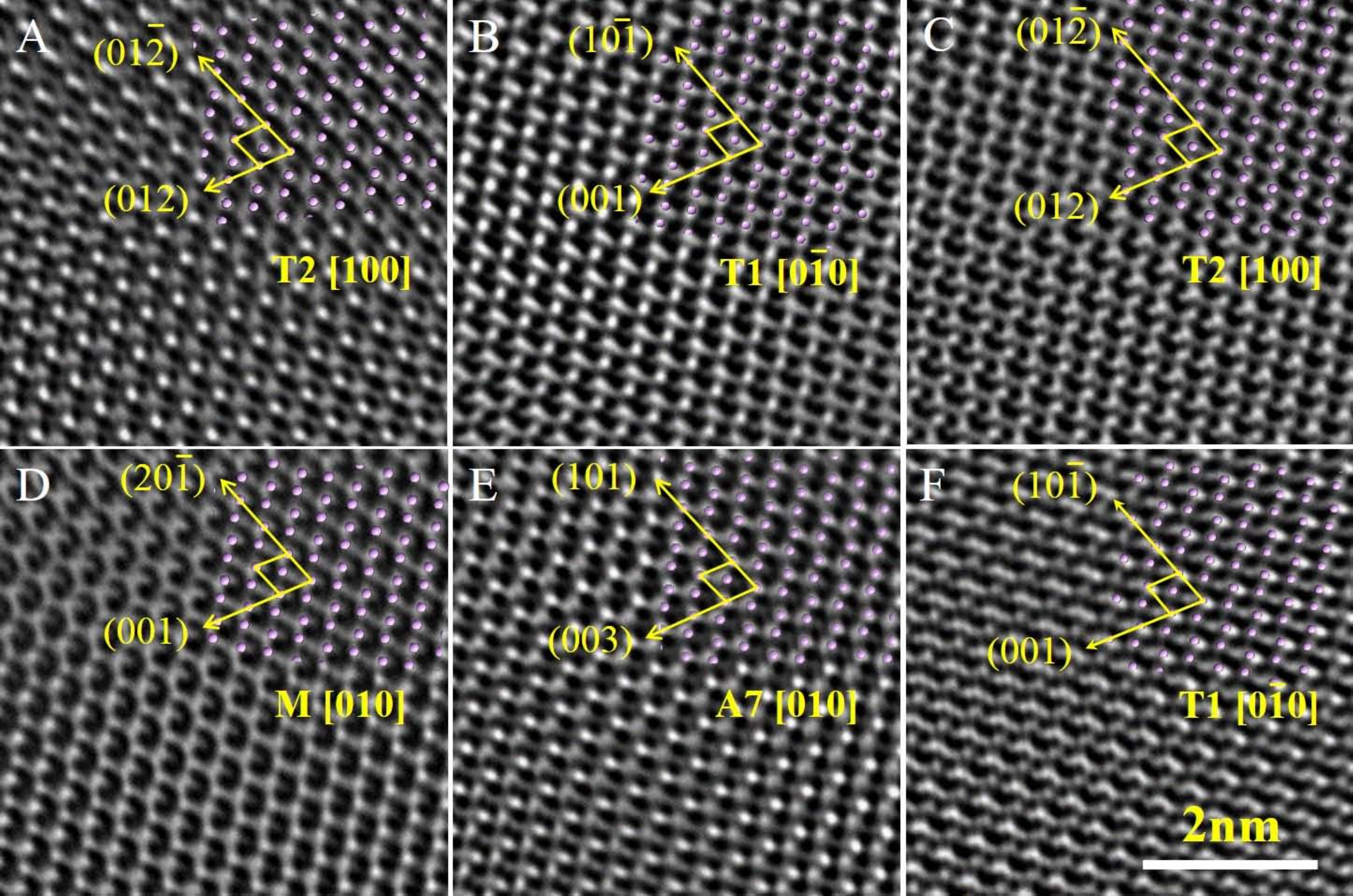}
\caption{\label{}Enlarged atomic-resolution electron micrographs for A-F areas in HRTEM image shown in Fig.1. Corresponding atomic arrangements for given Bi structures are projected on the micrographs.}
\end{figure}

The shifts of Bi atoms from the A7 structure are so small ( less than 0.1 \AA\ ) that HRTEM image and the corresponding SAED patterns alone cannot offer solid proof of structural polytypes. To account for the subtle lattice distortions observed in HRTEM measurement, we theoretically investigated a number of possible structural variations, in which atoms were shifted slightly from the equilibrium sites of the A7 structure without significant energy increase. Three new structures based on CALYPSO searches along with the known A7 structure and phase II (structure of Bi at pressure higher than 2.7 GPa) are shown in Fig. 3, where each structure is depicted with biatomic layers to highlight the structural similarities among them. These newly identified structures (Fig. 3b--d) are designated as M (monoclinic with space group C2/m), T1 (triclinic with space group P-1), and T2 (another triclinic structure with space group P1) phases, respectively. The symmetry groups, enthalpies, densities, and lattice parameters of these structures are listed in Table I. Unit cells and atomic positions are shown in Fig. S1 and Table SIII\cite{[]Supplemental24}, respectively.
\begin{figure}[h]
\centering
\includegraphics[scale=0.22]{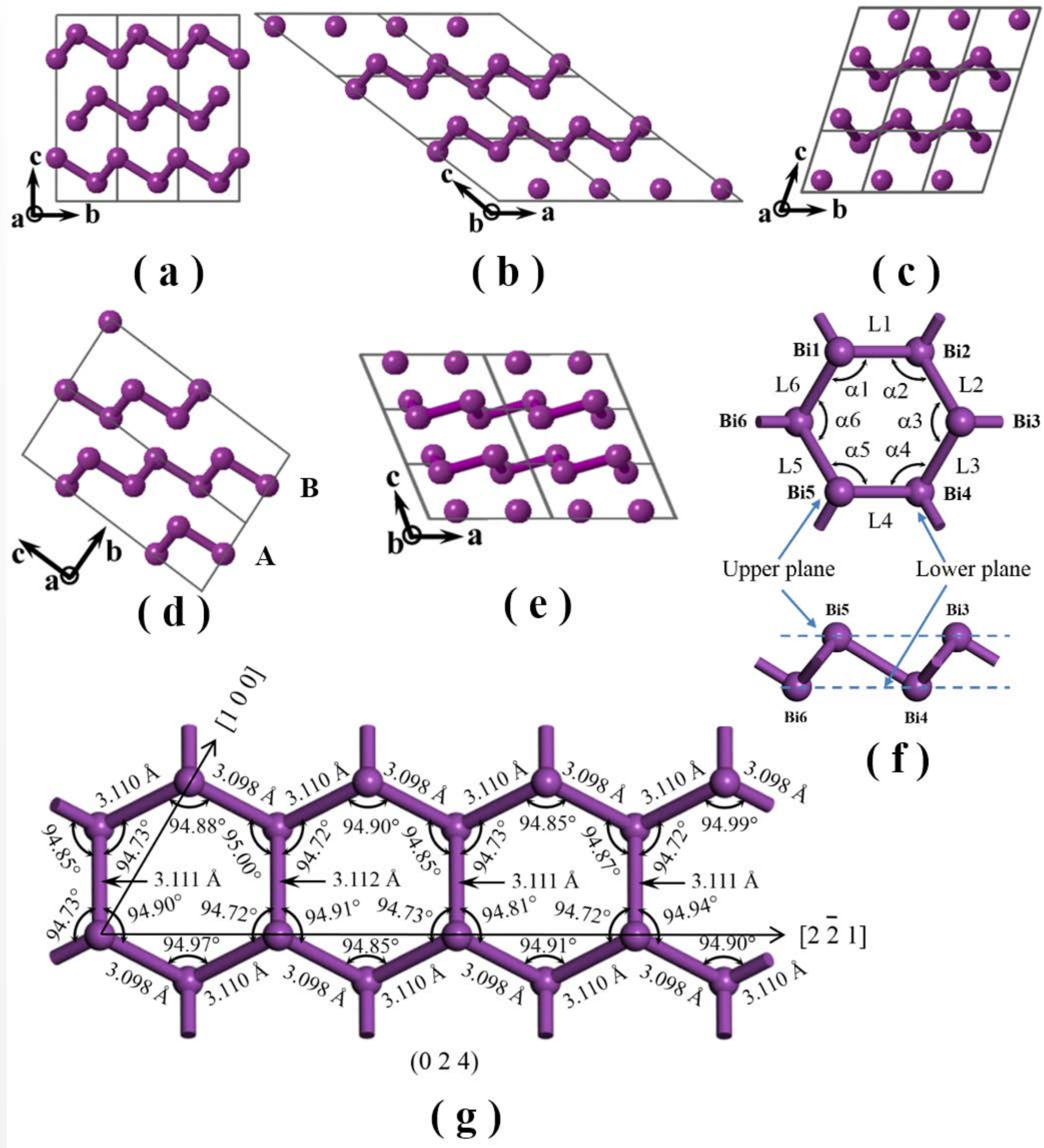}
\caption{\label{}Top views of possible Bi polytype crystal structures. (a)--(e) are views along the $\mit{a}$ or $\mit{b}$ axis for A7, M, T1, T2, and Bi-phase II, respectively. The double atomic layers are described by the aotms connevted by illustrative bonding sticks; (f) the typical chair-form hexatomic unit, and (g) the distorted chair-form hexatomic unit for the T2 structure.}
\end{figure}

The bismuth biatomic layer can be constructed by chair-formed hex-atomic rings as building blocks. In Fig. 3f, Bi atoms labelled as Bi1, Bi3, and Bi5 in the hex-atomic ring constitute the upper sub-layer, while those labelled as Bi2, Bi4, and Bi6 constitute the lower sub-layer. The Bi structures are constructed by stacking a series of biatomic layers. The biatomic layers in A7, M, T1, and Phase II structures are all parallel to the $\mit{ab}$ plane of the lattice. The A7 structure is comprised of $\cdots$ABCABC$\cdots$ stacked biatomic layers A, B, and C along the $\mit{c}$ axis, with each layer spanning one third of the lattice period. For M, T1, and phase II structures, each biatomic layer spans one lattice period along the $\mit{c}$ axis. T2 structure (Fig. 3d), however, possesses the most complex crystal structure: The repeating unit consists of four chair-formed hex-atomic rings with distinct bond angles and bond lengths, as detailed in Fig. 3g. Each biatomic layer can only be constituted by moving the units along the [100] and [2$\bar{2}$1] directions. These layers are parallel to the (024) plane of T2 structure, with an $\cdots$ABAB$\cdots$ stacking along the $\mit{b}$ axis.
\begin{table}[h]
\begin{ruledtabular}
\centering  
\tiny
\caption{Calculated symmetry groups, lattice parameters, enthalpies, densities, and double-atomic layer characteristics of the structural polytypes.}
\setlength{\tabcolsep}{1pt}
\begin{tabular}{cccccc}

 &Symmetry &lattice &Enthalpy &Density & layer structural\\
\vspace{1ex}
\raisebox{2ex}[0pt]{Structure}~ &group &parameter &(eV/atom) &(g/cm$^3$) &characteristic\\ \hline
\vspace{-1ex}
 & &a=b=4.5540 \AA\ & & &$\alpha$1=$\alpha$2=$\alpha$3=$\alpha$4\\
 \vspace{-0.3ex}
A7 &R-3M &c=11.8281 \AA\ &-3.8783 &9.82 &=$\alpha$5=$\alpha$6=94.54$^o$\\
\vspace{-1ex}
 &($D^{5}_{3d}$) &$\alpha$=$\beta$=90$^o$ & & &L1=L2=L3=L4\\
\vspace{1.5ex}
 & &$\gamma$=90$^o$ & & &=L5=L6=3.1 \AA\ \\

\vspace{-1ex}
 & &a=7.8873 \AA\ & & &$\alpha$1=$\alpha$2=$\alpha$4=$\alpha$5=94.55$^o$\\
\vspace{-0.3ex}
M &C2/m &b=4.5572 \AA\ &-3.8777 &9.75 &$\alpha$3=$\alpha$6=94.70$^o$\\
\vspace{-1ex}
 &($C^{3}_{2h}$) &c=6.5836 \AA\ & & &L1=L4=3.102 \AA\ \\
 \vspace{-1ex}
 & &$\alpha$=$\gamma$=90$^o$ & & &L2=L3=L5 \\
 \vspace{1.5ex}
 & & $\beta$=143$^o$ & & &=L6=3.098 \AA\ \\

\vspace{-1ex}
 & &a=4.5738 \AA\ & & &$\alpha$1=$\alpha$4=94.98$^o$\\
\vspace{-1ex}
 & &b=4.5786 \AA\ & & &$\alpha$2=$\alpha$5=94.78$^o$\\
\vspace{-0.3ex}
T1 &P--1 &c=4.8088 \AA\ &-3.8769 &9.47 &$\alpha$3=$\alpha$6=94.61$^o$\\
\vspace{-1ex}
 &($C^{1}_{i}$) &$\alpha$=90.31$^o$ & & &L1=L4=3.098 \AA\ \\
\vspace{-1ex}
 & &$\beta$=118.25$^o$ & & &L2=L5=3.123 \AA\ \\
 \vspace{1.5ex}
 & &$\gamma$=60.04$^o$ & & &L3=L6=3.106 \AA\ \\

\vspace{-1ex}
 & & & & &$\alpha$1=94.92$\pm$0.07$^o$\\
\vspace{-1ex}
 & &a=4.5767 \AA\ & & &$\alpha$2=94.92$\pm$0.08$^o$\\
\vspace{-1ex}
 & &b=4.7786 \AA\ & & &$\alpha$4=94.91$\pm$0.06$^o$\\
\vspace{-0.3ex}
T2 &P1 &c=13.3039 \AA\ &-3.8781 &9.56 &$\alpha$5=94.88$\pm$0.07$^o$\\
\vspace{-1ex}
 &($C^{1}_{1}$) &$\alpha$=86.07$^o$ & & &$\alpha$3=$\alpha$6=94.725$\pm$0.005$^o$ \\
\vspace{-1ex}
 & &$\beta$=90.02$^o$ & & &L1=L4=3.098 \AA\ \\
\vspace{-1ex}
 & &$\gamma$=90.02$^o$ & & &L2=L5=3.1115$\pm$0.005 \AA\ \\
\vspace{1.5ex}
 & & & & &L3=L6=3.110 \AA\ \\

\vspace{-1ex}
 & &a=7.0015 \AA\ & & &$\alpha$1=$\alpha$3=$\alpha$4=$\alpha$6=94.91$^o$\\
\vspace{-0.3ex}
Phase \uppercase\expandafter{\romannumeral2} &C2/m &b=6.2556 \AA\ &-3.8406 &10.07 &$\alpha$2=$\alpha$5=145.28$^o$\\
\vspace{-1ex}
 &($C^{3}_{2h}$) &c=3.3749 \AA\ & & &L1=L2=L4 \\
 \vspace{-1ex}
 & &$\alpha$=$\gamma$=90$^o$ & & &=L5=3.277 \AA\ \\
 \vspace{1.5ex}
 & & $\beta$=112.71$^o$ & & &L3=L6=3.199 \AA\ \\

\end{tabular}
\end{ruledtabular}
\end{table}

The chair-formed hex-atomic rings in these structures differ slightly in the bond angle and bond length. The corresponding angles ($\alpha$1--$\alpha$6) and lengths (L1$\--$L6) are thus introduced (see Fig. 3f), and listed in Table I to characterize each structure. The A7 structure has a unique chair conformation with identical bond angle of 94.54$^o$ and bond length of 3.1 \AA\ . For the newly identified Bi polytypes, the bond angle and length show gradually increased diversities, leading to a gradually reduced symmetry from M to T1, and to T2. However, we note that deviations in the bond angle, bond length, and enthalpy are only within 0.5$\%$, 0.8$\%$, and 0.04$\%$, respectively, for the Bi polytypes compared with the A7 structure.

Dynamic stability of the new structures was verified by the phonon spectrum calculations (Fig. S2, \cite{[]Supplemental24}), which show no imaginary frequencies in the Brillouin zone. Thermodynamic stability of these derivative structures was also examined, and the results are shown in Fig. 4. Enthalpy differences of the new structures relative to the A7 structure are less than 3 meV at pressures between 0 and 2.7 GPa. Such small differences \begin{figure}[h] \centering \includegraphics[scale=0.36]{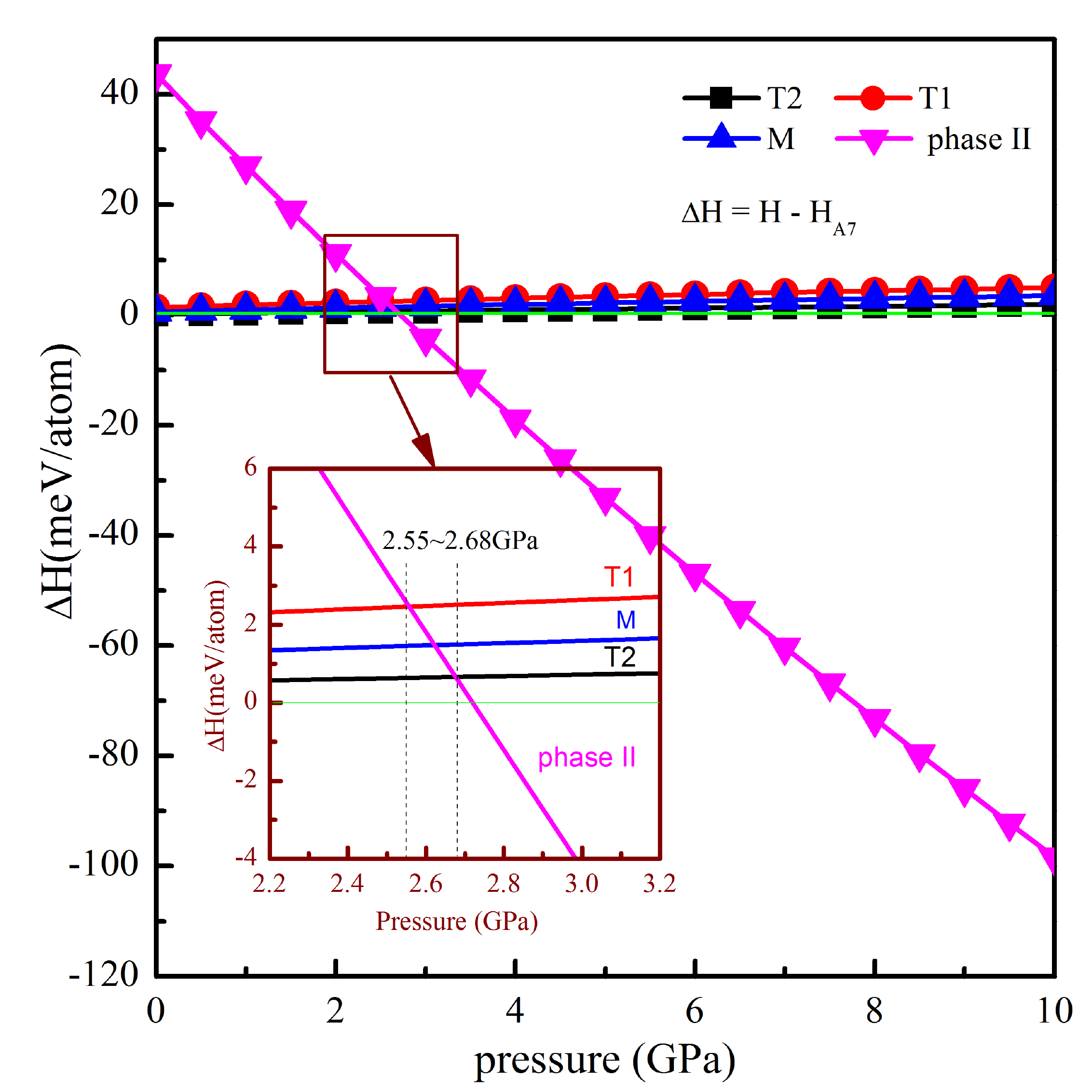} \caption{\label{}Enthalpy differences ($\Delta$H) of the predicted Bi polytype structures relative to A7. Inset shows the details of $\Delta$H at the pressure range from 2.2 to 3.2 GPa.} \end{figure}
suggest all these new structures are stable within this pressure range. At pressures higher than 2.7 GPa, our calculations show that the most stable structure is Bi-phase II. This is consistent with the known phase diagram of Bi \cite{[]Heremans8}.

The very small enthalpy differences among the A7, M, T1, and T2 structures suggest that very small energy perturbations, due to either phase transition from phase II on pressure release or magnetic treatments, may alter the structure of Bi and that these derivative polytypes can co-exist with the A7 structure. This may be the structural origin of the speculated ``mosaic block" or secondary structure in pure Bi \cite{[]Goetz18,Goetz19,Zwicky20,Zwicky21}. The density of calculated A7 structure is 9.80 g/cm$^3$, in agreement with the experimental value reported by Kapitza \cite{[]Kapitza17}. The densities of M, T1, and T2 structures are all slightly lower than that of A7, with the lowest being 9.47 g/cm$^3$ in T1 structure. Thus, the coexistence of various structural polytypes explains well the density variations measured by Goetz in Bi single crystals solidified under different magnetic field strength \cite{[]Zwicky21}.

Table SII summarizes d-spacings and interplanar angles measured from HRTEM for regions A--F of Fig. 2 in comparison with the calculated ones \cite{[]Supplemental24}. Simulated atomic positions viewed along the corresponding directions of the structural polytypes are superimposed on the HRTEM images in the upper right corners (Fig. 2). The excellent agreement between simulated and observed images suggest that region E corresponds to a projection along the [010] direction of the A7 structure, regions A and C to a projection along T2 [100], regions B and F to a projection along T1 [0$\bar{1}$0], and region D to a projection along M [010].

The structural polytypes can also be detected in XRD patterns (Fig. 5), which further confirm the coexistence of the derivative polytypes in bulk samples. XRD peaks generally match the rhombohedral A7 structure (PDF$\#$44-1246), but with additional weak peaks near the main peaks that cannot be explained
\begin{figure}[h]
\centering
\includegraphics[scale=0.38]{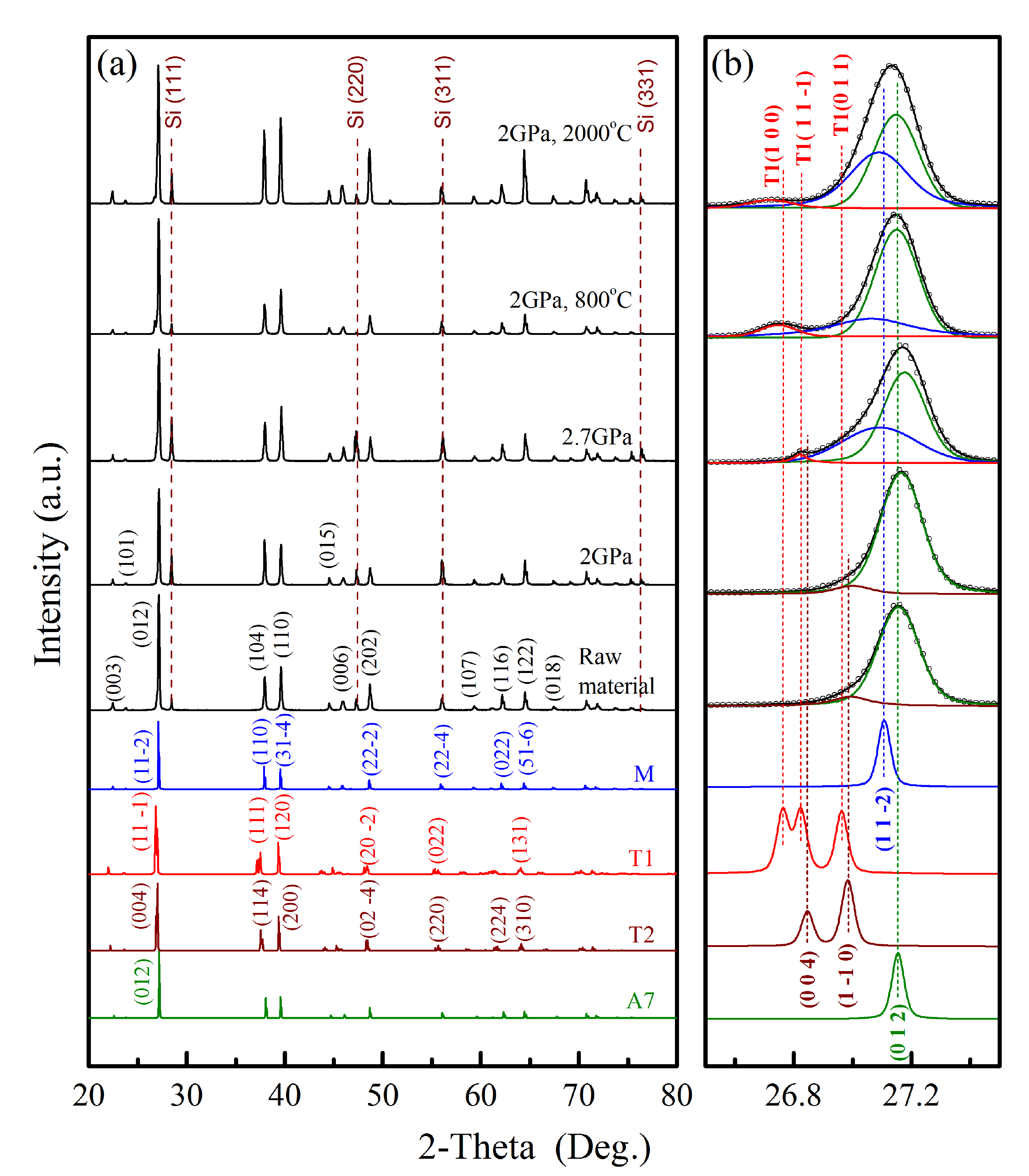}
\caption{\label{}Observed XRD patterns of various Bi samples as compared with those calculated new from the new structural polytypes. (a) Experimental XRD spectra corresponding to the Bi samples obtained under raw state, cold-pressing at 2GPa and 2.7GPa, HT-HP at 2GPa-1073 K and 2GPa-2273 K, the calculated XRD spectra of M, T1, T2 and A7 structures. (b) The enlarged spectra at the range of 26.5--27.5$^{\circ}$.}
\end{figure}
by the A7 structure alone. The calculated XRD patterns of the M, T1, T2, and A7 structures are also shown in Fig. 5 for comparison. For illustration purposes, we concentrate on the main peak in the XRD patterns, which are magnified in Fig. 5b for the two-theta range from 26.5$^o$ to 27.5$^o$. Clearly, there are one or more additional peaks other than the A7 (012) peak at 27.16$^o$ in all the samples. A small peak at 26.99$^o$ corresponding well to T2 (1$\bar{1}$1), is present in the raw material (untreated) and the sample cold-press to 2 GPa, suggesting that small amounts of the derivative polytype T2 phase coexisted in the samples. Two additinal peaks at 26.82$^o$ and 27.09$^o$ present in the the sample cold-press to 2.7 GPa (i.e., above the A7 $\leftrightarrow$ phase II boundary of 2.52 GPa) can be attributed to the T1 (11$\bar{1}$) and M (11$\bar{2}$), respectively. In the samples treated at 2 GPa/1073 K and 2 GPa/2273 K, the peaks at 26.75$^o$ and 27.09$^o$ may correspond to the diffraction lines of T1 (100) and M (11$\bar{2}$), respectively (Fig. 5b). Note that the last two samples underwent melting and recrystallization, a more intense new peak is present at a smaller angle.

Based on the experimental and theoretical investigations, we establish the structural diversity as an essential property for element Bi, in particular for samples treated at high pressures and high temperatures. Such diversity leads to the formation of secondary structures, which may be considered as derivative structures of the A7 phase. Coexisting derivative structural polytypes may be a general phenomenon. One example is the monoclinic diamond (M-diamond) recently reported in bulk nanotwinned diamond material \cite{[]Huang31}. In that case, phase transition from carbon onion nanoparticles to diamond and localized stress heterogeneity across diamond twin domains may be the cause. The exceptional thermal and mechanical properties observed in such multi-phase material suggest that coexisting structural polytypes may have interesting potentials in tailoring physical properties of materials.

\section{Acknowledgments}
\begin{acknowledgments}
This work was supported by the National Natural Science Foundation of China (51172197, 51121061, 51332005, 11025418 and 91022029), the Ministry of Science and Technology of China (2011CB808205 and 2010CB731605).
\end{acknowledgments}

\bibliography{references}

\end{document}



\title{Supplemental Material for ``Coexistence of multiple metastable polytypes in rhombohedral bismuth"}



\author{Yu Shu$^{1}$, Wentao Hu$^{1}$, Zhongyuan Liu$^{1}$, Guoyin Shen$^{2}$,
Zhisheng Zhao$^{1}$, \\Julong He$^{1}$, Yanbin Wang$^{3}$, Yongjun Tian$^{1}$, Dongli Yu$^{1*}$
}
\affiliation{$^{1}$State Key Laboratory of Metastable Materials Science and Technology, Yanshan University, Qinhuangdao, Hebei 066004, China\\
$^{2}$High Pressure Collaborative Access Team, Geophysical Laboratory, Carnegie Institution of Washington, Argonne, IL 604390, USA\\
$^{3}$Center for Advanced Radiation Sources, University of Chicago, Chicago, Illinois 60439, USA
}


\maketitle


\section{Research Methods}
\subsection{Experimental details}
Bi granules (Alfa Aesar, purity of 99.999$\%$) were compressed into cylinders (6 mm in diameter and 6 mm in height) and each was placed into a BN capsule with 8 mm outer diameter. High-pressure and high-temperature (HP-HT) experiments were performed using a China-type large volume cubic press with a maximum of 1400 tons on every WC anvil \cite{Liu23}. Pyrophyllite cubes with edge length of 49 mm served as both pressure medium and gasket. The samples were first compressed to required pressure, and then heated to the desired temperature for 1 hour by resistive heating with graphite tube heaters (outer diameter of 10 mm and inner diameter of 8 mm). The HP-HT conditions are presented in Table SI. For all the experiments, temperature was measured in situ with a type C thermocouple (W5/Re26). Pressure was estimated with previously obtained calibration curves based on the electrical resistance change\cite{Decker30} during the phase transitions of Bi at room temperature. After HP-HT treatment, the samples were quenched to room temperature at a cooling rate of $\sim$100K/s before the pressure was unloaded. Bulk samples recovered at ambient condition had diameters of approximately 5.5 mm and heights of 5.3--5.7 mm. Both the raw granules and the bulk samples after HP-HT treatment were ground in an agate mortar with a pestle, for XRD (Rigaku D/MAX-PC/2500, operated at 40 kV and 200 mA with a Cu-K$\alpha$ target) and transmission electron microscopy (JEM-2010 with an accelerating voltage of 200 kV). The HR-TEM was carefully calibrated with silicon at the selected magnification. The change in microstructure introduced due to grinding is expected to be the same for all samples; thus the difference observed in TEM and XRD analyses are considered due to HPHT treatment.
\subsection{First-Principles Calculations}
The software package Crystal structure AnaLYsis by Particle Swarm Optimization (CALYPSO)\cite{Ma25} was employed to search possible low-energy structures for Bi. We examined thousands of candidate structures with simulation cell sizes ranging from 2 to 12 atoms, at pressures of 0--6 GPa. Structural relaxations were performed using the density functional theory (DFT) within the generalized gradient approximation (GGA) implemented in the Vienna Ab Initio Simulation Package (VASP). Projector augmented-wave pseudopotential with 6s$^{2}$6p$^{3}$ electrons (as is the case of the valence for Bi) was adopted. The energy cutoff was 300 eV, and appropriate Monkhorst-Pack k meshes were chosen to ensure that enthalpy calculations converged to better than 0.01 meV per atom for each structure. Phonon frequencies were generated using the Cambridge Serial Total Energy Package (CASTEP) with finite displacement theory. Conventional DFT schemes that apply to the GGA usually overestimate lattice constant parameters\cite{Philipp29}. In order to make direct comparison between calculated lattice parameters and the experimentally observed values, we adjusted hyrostatic pressure in computing the lattice parameters of the A7 structure, until they match the parameters for the rhombohedral structure (PDF$\#$44-1246). This hyrostatic pressure was identified for 1.2 GPa and then applied to all other possible structures in first-principles calculations. Finally, XRD patterns of these structures were simulated by Reflex tools package of the Materials Studio (MS) software, to help interpreting the experimentally observed XRD patterns.

\bibliography{PRL-references}


\renewcommand*{\tablename}{Table S}

\begin{table}[p]
\centering  
\caption{Details of the HP-HT treatment conditions}
\begin{tabular}{cccc}  
\hline
\hline
\vspace{-1.5ex}  & & &Sample state under\\
\raisebox{1.8ex}[0pt]{Sample No.}~~ &\raisebox{1.8ex}[0pt]{Pressure (GPa)}~~ &\raisebox{1.8ex}[0pt]{Temperature (K)}~~ &treatment conditions \\ \hline 

1 &2$\pm$0.1 &303$\pm$5 &Phase \uppercase\expandafter{\romannumeral1} \\  
2 &2.7$\pm$0.1 &303$\pm$5 &Phase \uppercase\expandafter{\romannumeral2} \\
3 &2$\pm$0.1 &1073$\pm$10 &Liquid \\
4 &2$\pm$0.1 &2273$\pm$30 &Liquid \\ \hline
\hline
\end{tabular}
\end{table}

\begin{table}[p]
\centering  
\caption{ Correspondence between the measured data for different areas in HR-TEM and corresponding d-spacings and angles in predicted Bi structures}
\begin{tabular}{cccccc}  
\hline
\hline

\multicolumn{3}{c}{} &\multicolumn{3}{c}{Corresponding d-spacings and angles in} \\
\multicolumn{3}{c}{ \raisebox{1.8ex}[0pt]{Measuring data in HR-TEM image} } &\multicolumn{3}{c}{predicted Bi structures} \\ \hline 
\vspace{-1.5ex}
Area in HR- &d-spacing( \AA\ ) &Angle( $^{\circ}$ ) &Structure and & & \\
TEM image &Deviation:$\pm$0.03~ &Deviation:$\pm$0.05 &zone axis &\raisebox{1.8ex}[0pt]{~d-spacing( \AA\ )} &\raisebox{1.8ex}[0pt]{~Angle( $^{\circ}$) } \\ \hline

\vspace{-1.5ex}
 &$d_{1}$=3.99 & &T2-Bi &$d_{(012)}$=4.0 & \\
\raisebox{1.8ex}[0pt]{A} &$d_{2}$=3.76 &\raisebox{1.8ex}[0pt]{71.38} &[100] &$d_{(01 {\bar{2}})}$=3.7453 &\raisebox{1.8ex}[0pt]{71.34} \\

\vspace{-1.5ex}
 &$d_{1}$=3.97 & &T1-Bi &$d_{(001)}$=4.0323 & \\
\raisebox{1.8ex}[0pt]{B} &$d_{2}$=3.76 &\raisebox{1.8ex}[0pt]{71.54} &[0$\bar{1}$0] &$d_{(10{\bar{1}})}$=3.7593 &\raisebox{1.8ex}[0pt]{71.50} \\

\vspace{-1.5ex}
 &$d_{1}$=3.97 & &T2-Bi &$d_{(012)}$=4.0 & \\
\raisebox{1.8ex}[0pt]{C} &$d_{2}$=3.76 &\raisebox{1.8ex}[0pt]{71.36} &[100] &$d_{(01 {\bar{2}})}$=3.7453 &\raisebox{1.8ex}[0pt]{71.34} \\

\vspace{-1.5ex}
 &$d_{1}$=3.95 & &M-Bi &$d_{(001)}$=3.9683 & \\
\raisebox{1.8ex}[0pt]{D} &$d_{2}$=3.76 &\raisebox{1.8ex}[0pt]{71.67} &[010] &$d_{(20 {\bar{1}})}$=3.7453 &\raisebox{1.8ex}[0pt]{71.64} \\

\vspace{-1.5ex}
 &$d_{1}$=3.97 & &A7-Bi &$d_{(003)}$=3.9370 & \\
\raisebox{1.8ex}[0pt]{E} &$d_{2}$=3.74 &\raisebox{1.8ex}[0pt]{71.52} &[010] &$d_{(101)}$=3.7453 &\raisebox{1.8ex}[0pt]{71.56} \\

\vspace{-1.5ex}
 &$d_{1}$=3.97 & &T1-Bi &$d_{(001)}$=4.0323 & \\
\raisebox{1.8ex}[0pt]{F} &$d_{2}$=3.76 &\raisebox{1.8ex}[0pt]{71.43} &[0$\bar{1}$0] &$d_{(10{\bar{1}})}$=3.7593 &\raisebox{1.8ex}[0pt]{71.50} \\

\hline
\hline

\end{tabular}
\end{table}

\begin{table}[p]
\centering  
\caption{Atomic positions in new structures of Bi}
\begin{tabular}{cccccc}  
\hline
\hline
\vspace{-1.5ex}
 & & & &Atomic Positions &\\
\raisebox{1.8ex}[0pt]{Bi structures}~~ &\raisebox{1.8ex}[0pt]{Atoms}~~ &\raisebox{1.8ex}[0pt]{Wyckoff} &x/a &y/b &z/c \\ \hline 

M &Bi1 &4i &1.029(0) &0 &0.793(1) \\
T1 &Bi1 &2i &0.965(5) &0.269(6) &0.202(8) \\

\vspace{-1.5ex}
 &Bi1 &1a &0.125(0) &0.971(2) &0.991(7) \\
\vspace{-1.5ex}
 &Bi2 &1a &0.624(6) &0.528(2) &0.508(5) \\
\vspace{-1.5ex}
 &Bi3 &1a &0.124(8) &0.029(1) &0.758(4) \\
\vspace{-1.5ex}
T2 &Bi4 &1a &0.625(1) &0.470(7) &0.741(8) \\
\vspace{-1.5ex}
 &Bi5 &1a &0.124(3) &0.028(7) &0.258(3) \\
 \vspace{-1.5ex}
 &Bi6 &1a &0.624(6) &0.470(4) &0.241(9) \\
 \vspace{-1.5ex}
 &Bi7 &1a &0.124(9) &0.969(9) &0.491(7) \\
 &Bi8 &1a &0.624(6) &0.529(5) &0.008(5) \\  \hline
\hline

\end{tabular}
\end{table}

\renewcommand*{\figurename}{Figure S}
\begin{figure*}[p]
\centering
\includegraphics[scale=0.3]{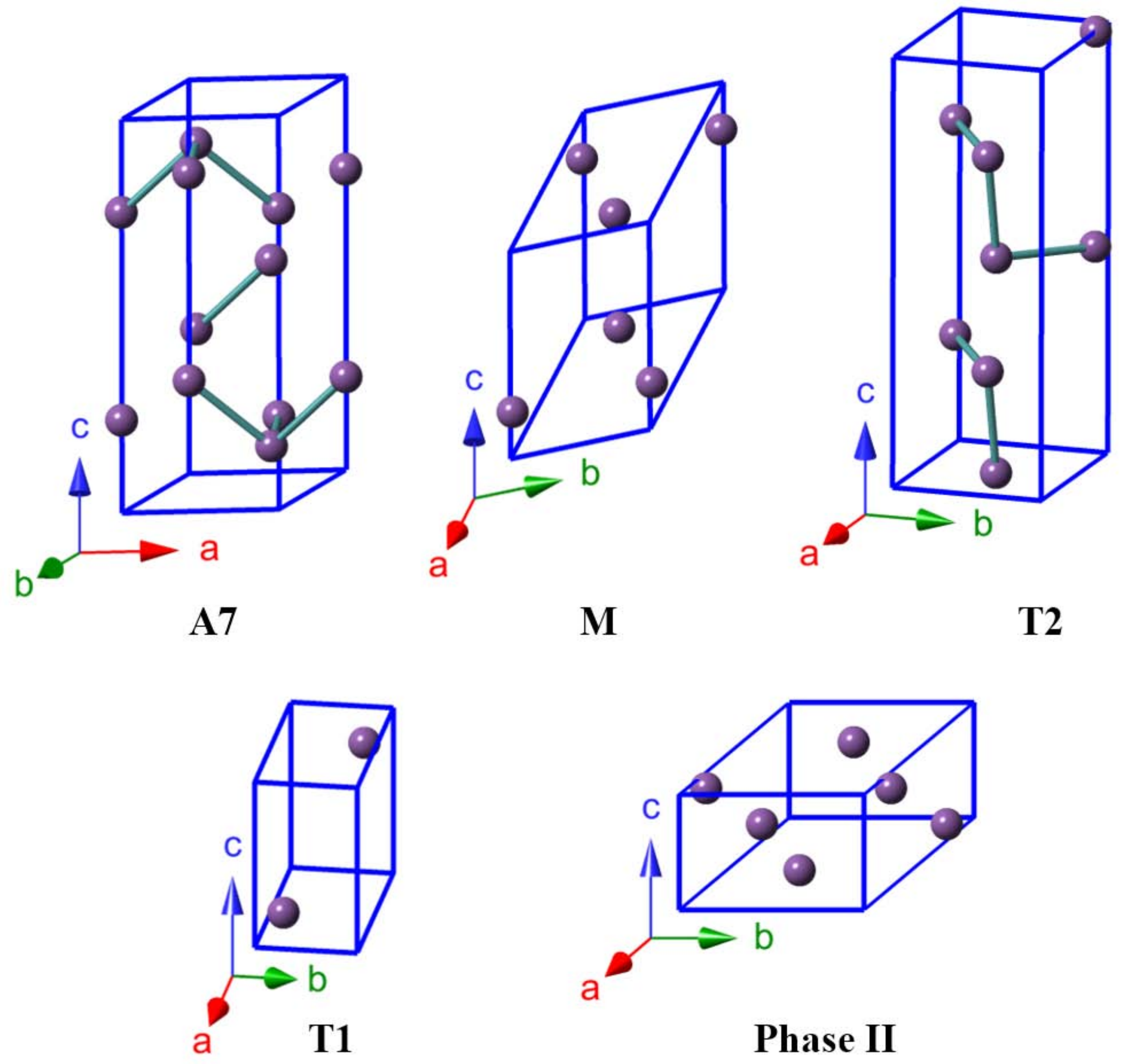}
\caption{\label{}Unit cells of the Bi crystal structures.}
\end{figure*}

\begin{figure*}[p]
\centering
\includegraphics[scale=0.5]{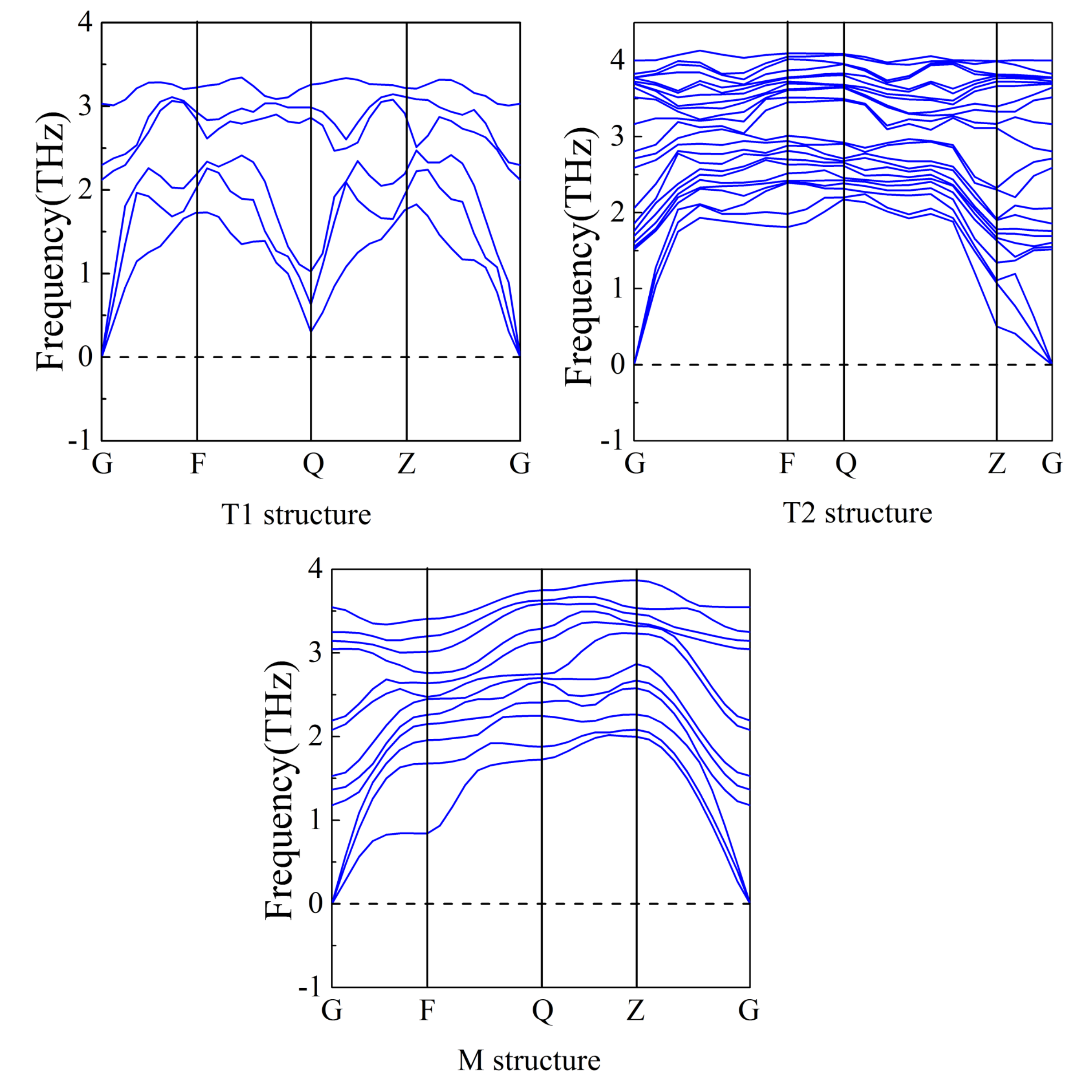}
\caption{\label{}Phonon spectra of new Bi structures.}
\end{figure*}